\documentstyle[12pt]{article}
\def\nonu{\nonumber}
\def\br{\begin{eqnarray}}
\def\er{\end{eqnarray}}
\def\be{\begin{equation}}
\def\ee{\end{equation}}

\def\({\left(}
\def\){\right)}

\relax

\def\Tr{\mathop{\rm Tr}}

\newcommand{\sbr}[2]{\left\lbrack\,{#1}\, ,\,{#2}\,\right\rbrack}

\newcommand\ket[1]{\mid \, {#1} \, \rangle}

\def\a{\alpha}

\def\b{\beta}

\def\d{\delta}

\def\vareps{\varepsilon}

\def\h{{1\over 2}}
\def\l{\lambda}

\def\pa{\partial}
\def\pr{\prime}

\def\ra{\rightarrow}
\def\s{\sigma}

\def\tp0{\Theta_{+}^{(0)}}
\def\tm0{\Theta_{-}^{(0)}}

\def\u2{\mid u\mid^2}

\def\vp{\varphi}

\def\ck{{\cal K}}
\def\cl{{\cal L}}

\def\f#1#2#3 {f^{#1#2}_{#3}}

\def\win1{{\sf w_{1+\infty}}}

\def\Win1{{\sf W_{1+\infty}}}

\def\rlx{\relax\leavevmode}
\def\inbar{\vrule height1.5ex width.4pt depth0pt}
\def\IZ{\rlx\hbox{\sf Z\kern-.4em Z}}
\def\IR{\rlx\hbox{\rm I\kern-.18em R}}
\def\IC{\rlx\hbox{\,$\inbar\kern-.3em{\rm C}$}}
\def\IN{\rlx\hbox{\rm I\kern-.18em N}}
\def\IO{\rlx\hbox{\,$\inbar\kern-.3em{\rm O}$}}
\def\IP{\rlx\hbox{\rm I\kern-.18em P}}
\def\IQ{\rlx\hbox{\,$\inbar\kern-.3em{\rm Q}$}}
\def\IF{\rlx\hbox{\rm I\kern-.18em F}}
\def\IG{\rlx\hbox{\,$\inbar\kern-.3em{\rm G}$}}
\def\IH{\rlx\hbox{\rm I\kern-.18em H}}
\def\II{\rlx\hbox{\rm I\kern-.18em I}}
\def\IK{\rlx\hbox{\rm I\kern-.18em K}}
\def\IL{\rlx\hbox{\rm I\kern-.18em L}}
\def\one{\hbox{{1}\kern-.25em\hbox{l}}}
\def\0#1{\relax\ifmmode\mathaccent''7017{#1}%
B        \else\accent23#1\relax\fi}

\def\AoP#1#2#3{{\sl Ann. of Phys.} {\bf #1} (#2) #3}

\begin{document}

\begin{titlepage}
\vspace*{-1cm}

\noindent
October, 2002 \hfill{IFT-P.077/02}\\
hep-th/0210154 \hfill{LPTHE-02.52}

\vskip 3cm

\vspace{.2in}
\begin{center}
{\large\bf Integrability and Conformal Symmetry in Higher
  Dimensions: A Model with 
  Exact Hopfion Solutions}
\end{center}

\vspace{.5cm}

\begin{center}
O. Babelon~$^1$ and L. A. Ferreira~$^2$

\vspace{.5 in}
\small

\par \vskip .2in \noindent
$^{(1)}$~Laboratoire de Physique Theorique et Hautes Energies\\
Universit\'es Paris VI - Paris VII\\ 
Bo\^\i te 126, Tour 16, 1$^{\rm er}$ \'etage\\ 
4 Place Jussieu, F-75252, Paris CEDEX 05, France

\par \vskip .2in \noindent
$^{(2)}$~Instituto de F\'\i sica Te\'orica - IFT/UNESP\\
Rua Pamplona 145\\
01405-900  S\~ao Paulo-SP, Brazil\\

\normalsize
\end{center}

\vspace{.5in}

\begin{abstract}
We use ideas on integrability in higher dimensions to define Lorentz
invariant  field theories with an infinite number of local
conserved currents. The models considered have a two dimensional
target space. Requiring the existence of Lagrangean and the stability
of static solutions 
singles out a class of models which have an additional conformal
symmetry. That is used to 
explain the existence of an ansatz leading to solutions with non trivial
Hopf charges. 
  
\end{abstract}
\end{titlepage}

\section{Introduction}

We consider in this paper Lorentz invariant local field theories in a
space-time of $d+1$ dimensions, with an infinite number of local
conserved currents.  In order
to construct such models we use the generalized zero curvature
conditions in any dimension proposed in \cite{AFSG97}. The basic
ingredient of that approach is to take the potentials of the zero
curvature to lie in non-semisimple Lie algebras with an infinite
dimensional abelian ideal. Restricting ourselves to theories with 
two dimensional target space,  we obtain those algebraic structures
using a 
$sl(2)$ algebra realized in terms of differential operators on two parameters
(Schwinger construction) and taking the ideal as the 
space  of functions of these parameters. 

The equations of motion
of the theories we obtain from the generalized zero curvature are of
the form   
\be
\partial^{\mu}{\cal K}_{\mu}=0 \qquad\qquad
\partial^{\mu} u {\cal K}_{\mu}=0\qquad \qquad
 \partial^{\mu} u^* {\cal K}_{\mu} - 
 \partial^{\mu} u {\cal K}_{\mu}^* =0
\label{eqofmotion}
\ee
where $u$ is a complex scalar field parametrizing our two-dimensional
target space, and ${\cal K}_{\mu}$ is a complex function of the fields
$u$ and $u^*$ and their  derivatives. These theories have an infinite
number of local 
conserved currents given by
$$
J_\mu =   {\cal K}_\mu {\delta G \over \delta u} -  
{\cal K}^*_\mu {\delta G \over \delta u*} 
$$
with $G$ being any functional of the fields $u$ and $u^*$ but not of
their derivatives. 
 
Since eqs.(\ref{eqofmotion}) are five real equations for two real fields, we
have a 
constrained system. However, there is a class of functionals  
${\cal K}_{\mu}$ that automatically satisfies the constraints. They are
given by 
\be
{\cal K}_{\mu}= {\cal F} h_{\mu\nu}\partial^{\nu} u,\qquad \qquad 
h_{\mu\nu}=\pa_{\mu}u\, \pa_{\nu} u^* - \pa_{\mu}u^*\, \pa_{\nu}u
\label{hmunudef}
\ee
where ${\cal F}$ is a real function of $u$, $u^*$, and their first order
derivatives. The above quantity solves automaticaly the last two
equations in (\ref{eqofmotion}).  

We show that the only  unconstrained theories
that can be derived
from an action principle are those where the Lagrangean is a function of
$h^2/f^2$ only, with $f$ being a real function of $u$ and $u^*$ and $h^2 =
h_{\mu\nu}h^{\mu\nu}$.   
For
these theories the currents above are Noether currents corresponding to the
invariance of the action under area preserving diffeomorphisms on
target space equipped with an area form $du\wedge du^*/f$.  

In order to have static solution we have to overcome Derrick's
theorem. Since $h^2/f^2$ scales as $\l^{-4}\; h^2/f^2$ under the
scaling transfomation $x^{\mu} \ra \l x^{\mu}$, we consider the
Lagrangeans 
\be
{\cal L} = \( \frac{h^2}{2f^2}\)^{d/4}
\label{scaleinvlag}
\ee
which give scale invariant static energy in $d$ space
dimensions. 

Such scale invariance has led us to study more closely  the  symmetry
group of the equations of motion. We found that the space-time and
target space symmetries always commute. We found that in the static
case, the equations of motion have the full conformal symmetry group
$O(d+1,1)$. In the non-static case only the scale transformation is a
symmetry of the equations of motion, besides the usual Poincar\'e
group. However, that scale transfomation is not a symmetry of the
action. In target space, we
found that besides the area preserving diffeomorphisms there is one
extra symmetry of the full equations of motion, which however is not a
symmetry of the Lagrangean either. We can however combine that with
the space-time scale transformation to build a true symmetry of the
action and then obtain an extra Noether current.   

Applying S. Lie's ideas about symmetries of differential equations to
the conformal symmetry group, we are immediatly led to construct
educated ansatz for solutions of the equations of motion. For the
theory (\ref{scaleinvlag}) with $d=3$, we used the
$O(4,1)$ conformal symmetry to introduced two cyclic variables
(angles) and then reduce the static equations of motion to an ordinary
differential equation for a profile function. We explain in this way
the origin of the ansatz used in \cite{AFZ992} to construct the exact
hopfion solutions with non-trivial Hopf charges. 

The charge associated to the extra Noether current, coming from the
 mixing of target space and 
 space-time scale transformations, is not conserved due to boundary
 terms. Its rate 
 of change in time is in fact the static energy. We have evaluated the
 flux of that current for the case of the $3$-dimensional hopfion
 solutions and found that the leaking of the charge takes place 
 along  the $z$-axis, where that flux is singular. This implies that
 the hopfion solutions carry a line structure with them associated to
 such line of singularity of the Noether current. This may be
 relevant in the scatterring process of those solutions.

\section{Zero curvature on loop spaces}
\label{sec:zerocurvature}
One of the basic ingredients of  two-dimensional integrable models
is the so-called Lax-Zakharov-Shabat zero curvature condition 
\be
\partial_\mu B_\nu - \partial_\nu B_\mu + [B_\mu, B_\nu] = 0 
\qquad \qquad \mu , \nu = 0,1
\label{lzs}
\ee
It is the starting point for several methods for constructing exact
solutions, like the dressing, inverse scattering and Riemmann-Hilbert
methods. In addition, in 
$1+1$ dimensions the relation (\ref{lzs}) is a conservation law, since it
is a sufficient condition for the path ordered integral 
\be
W(\Gamma) = P_\Gamma \exp \int_\Gamma  B_\mu dx^\mu  
\label{wilson}
\ee
to be independent of the path $\Gamma$, as long as the end points are
kept fixed. Consequently, on a closed loop one  has that $W$ is a
constant and so can be set  to
unit. Therefore, by choosing appropriate boundary conditions at space
infinity (or taking space to be a circle and so space-time a cylinder)
one gets that the quantities 
\be
Q_N \equiv \Tr\( P  \exp \int_{\rm space}  B_x dx  \)^N
\label{conscharges}
\ee
are conserved in time. These are the conservation laws  responsible for the
integrability of 
a two dimensional theory possessing the zero curvature
representation (\ref{lzs}). 

Under these considerations it is then natural to try to generalize the
Lax-Zakharov-Shabat  equation (\ref{lzs}) to theories defined on  
a  space-time of dimension $d+1$ as the sufficient condition for
ordered integrals of a generalized connection on a $d$ dimensional surface to
be surface independent. Similar reasonings and appropriate boundary
conditions should then lead to conservation laws. That is exactly the
idea put foward in \cite{AFSG97}.  The equivalent of (\ref{wilson}),
in $d+1$ dimensions, would be 
a $d$ dimensional surface ordered integral of a  rank $d$ 
antisymmetric tensor $B_{\mu_1 \ldots \mu_d}$. 

An important issue in the implementation of such ideas is the ordering of
the integration on the hypersurfaces. Given a $d$-dimensional surface
$\Sigma$ with border $\pa \Sigma$, we  choose a base point $x_0$
in $\pa \Sigma$ and then scan $\Sigma$ with $(d-1)$-dimensional closed
surfaces containing $x_0$. Such closed surfaces are ordered and
labeled by a parameter $\sigma_d$ such that, for instance,
$\sigma_d=0$ correspond to the infinitesimally small surface around
$x_0$, and $\sigma_d=2\pi$ to $\pa \Sigma$. Then each one of these
surfaces are scanned 
in their turn by $(d-2)$-dimensional closed surfaces with base point
$x_0$ too, and labeled by a parameter $\sigma_{d-1}$. The process goes
on until we reach two dimensional surfaces 
which are scanned by ordinary closed loops starting and ending at
$x_0$.

 These ideas find their best
formulation, in fact, in generalized loop 
space.  Given a $d+1$ dimensional space time $M$, one considers the
space $\Omega^{(d-1)}\(M,x_0\)$ of $d-1$ closed surfaces in $M$ with a
fixed base point $x_0$. A path in $\Omega^{(d-1)}\(M,x_0\)$ is a $d$
dimensional surface in $M$. Therefore, the conserved charges can be
obtained from path ordered integrals in the generalized loop space, in 
a similar way as charges in $1+1$ dimensions are obtained from path
ordered integral in space-time (see (\ref{conscharges})). Indeed, the
two approaches become the same in  $1+1$ dimensions, since for $d=1$
the loop space coincides with space-time. In addition, as we show
below, the rank $d$ antisymmetric tensor $B_{\mu_1 \ldots \mu_d}$ can
be used to construct a one-form connection in generalized loop space
by integrating it on a generalized loop ($(d-1)$-dimensional closed surface).

Notice that in the scanning just described we have a nested structure
of generalized loop spaces, since the closed $(d-n)$-dimensional
surfaces have base point $x_0$ and so belong to
$\Omega^{(d-n)}\(M,x_0\)$. We can then introduce, in the space-time
$M$, a set of
antisymmetric tensors of rank varying from $1$  to $d$ and use them to
construct one-form connections on each one of those loop
spaces. 

The zero curvature conditions for such loop space connections
usually leads to highly non-local expressions in space-time. In order
to make things as local as possible, it was found in \cite{AFSG97}
that it is advantageous to parallel
transport quantities in a given loop space  using the connections in
the lower loop spaces. Of 
course we do not want things to depend upon the way we scan the
surfaces. It then follows that all those connections on loop spaces
will have to be flat to guarantee the scanning independence. Therefore,
we have a nested structure of zero curvatures involving those
antisymmetric tensors. However, only the loop space connection
associated to the the rank $d$ tensor leads to conserved charges in
$d+1$ dimensions. We refer to \cite{AFSG97} for a more  detailed
explanation of the construction of such connections.

In the applications of the present paper we only consider the highest
and lowest tensors, i.e. a 
rank $d$ antisymmetric tensor $B_{\mu_1 \ldots \mu_d}$, and a vector 
$A_{\mu}$. In the scanning of  the $d$-dimensional surface $\Sigma_d$,
each one of its points belong to just one $(d-1)$-dimensional surface
$\Sigma_{d-1}$. Similarly, that point of $\Sigma_d$ lying in
$\Sigma_{d-1}$ belongs to only one $(d-2)$-dimensional closed surface
$\Sigma_{d-2}$ scanning $\Sigma_{d-1}$. Repeating this procedure we
see that each point of $\Sigma_d$ belongs to just one $1$-dimensional
loop of the scanning. We then use that loop and $A_{\mu}$ to parallel
transport $B_{\mu_1 \ldots \mu_d}$ at that point of $\Sigma_d$ back to
the base point $x_0$.

The surface $\Sigma_d$ is a path in the
generalized loop space $\Omega^{(d-1)}$. Therefore, the ordered
integral of the rank $d$ antisymmetric tensor $B_{\mu_1 \ldots \mu_d}$
over $\Sigma_d$, 
can be seen as a path ordered integral in $\Omega^{(d-1)}$ of a
one-form connection. We define that one-form on the generalized loop space
$\Omega^{(d-1)}$ by the formula 
\be
{\cal A} = \int_0^{2\pi} d\s_1\ldots d\s_{d-1} \; W^{-1}\, B_{\mu_1 \ldots
  \mu_d}\, W \;  
 {dx^{\mu_1} \over d \s_1}\ldots {dx^{\mu_{d-1}} \over d \s_{d-1}}
 \delta x^{\mu_d} 
\label{loopconnection}
\ee
where $\s_i$, $i=1,\ldots d-1$, parametrize the closed $d-1$
dimensional surface in $M$, according to the scanning described above. $W$ is
built as the path ordered 
integral of the one-form connection $A_{\mu}$, in a similar way as in
(\ref{wilson}). However, in order for things to be 
independent of the scanning we take the connection $A_{\mu}$ to be
flat. It then follows 
that $W$ is path independent, and so it is uniquely defined on every
point of $\Sigma_d$ once its value in $x_0$ 
is given. Therefore, the quantity $W^{-1}\, B_{\mu_1 \ldots
  \mu_d}\, W $, is local in $\Sigma_d$. The variation  $\delta x^{\mu} $
corresponds to 
infinitesimal changes in the closed $(d-1)$-dimensional surface
$\Sigma_{d-1}$, and it is in fact what 
characterizes ${\cal A}$ as a one-form in $\Omega^{(d-1)}$. Therefore,
the condition for the  ordered integral on a $d$ dimensional surface  
\be
{\cal P} \exp \(\int_{\Sigma_d} d\s_1\ldots d\s_{d} \; W^{-1}\,
B_{\mu_1 \ldots   \mu_d}\, W \;  
 {dx^{\mu_1} \over d \s_1}\ldots {dx^{\mu_{d}} \over d \s_{d}}\)
\ee
to be surface independent, is that the one-form ${\cal A}$ should be
flat, i.e.
\be
{\cal C} = \d {\cal A} + {\cal A} \wedge {\cal A} = 0
\label{genzc}
\ee
That is the generalization of the Lax-Zakharov-Shabat equation
(\ref{lzs}) proposed in  \cite{AFSG97}, for theories defined on a
space-time $M$ of $d+1$ dimensions. Under appropriate boundary
conditions it leads to the conserved charges
\be
\Tr\( {\cal P} \exp \(\int_{\rm space} d\s_1\ldots d\s_{d} \; W^{-1}\,
B_{\mu_1 \ldots   \mu_d}\, W \;  
 {dx^{\mu_1} \over d \s_1}\ldots {dx^{\mu_{d}} \over d \s_{d}}\)\)^N
\label{gencharges}
\ee
Notice that for $d=1$, one recovers the usual two-dimensional
Lax-Zakharov-Shabat  zero curvature condition. 

Clearly, the condition (\ref{genzc})
is still non-local in $M$, since ${\cal A}$  is a multidimensional integral
there. 
To get zero curvature representations of the equations of motion of
local field theories in the space-time $M$, we make special choice for 
$B_{\mu_1 \ldots   \mu_d}$ and $A_{\mu}$. It was shown in
(\ref{lzs})  that if one takes $A_{\mu}$ to live on a non-semisimple
Lie algebra and $B_{\mu_1 \ldots   \mu_d}$ to belong to the
corresponding abelian ideal then the local equations
\be
d B + A\wedge B = 0 \qquad\qquad  F = d A + A\wedge A =0
\label{localzc}
\ee
are sufficient conditions for the vanishing of the curvature 
(\ref{genzc})  in loop space. The second condition in
(\ref{localzc}) is imposed  to have things independent of the
way one scans the integrated surfaces. It implies that $A_{\mu}$ is
obtained from $W$ as a pure gauge connection, i.e.
\be
A_{\mu} = - \pa_{\mu} W \, W^{-1}
\label{puregauge}
\ee
We shall then take (\ref{localzc}) as our generalized zero curvature
conditions.  Introducing the dual of $B$
\be
{\tilde B}_{\mu} \equiv \frac{1}{d} \epsilon_{\mu \mu_1 \ldots \mu_d}
B^{\mu_1 \ldots   \mu_d} 
\label{dualb}
\ee
one gets that the first relation in (\ref{localzc}) becomes
\be
D^{\mu} {\tilde B}_{\mu} = \pa^{\mu} {\tilde B}_{\mu} + 
\sbr{A^{\mu}}{{\tilde B}_{\mu}} = 0
\label{dualeqforb}
\ee
As a consequence of such relation it follows that the conserved
charges (\ref{gencharges}) are now obtained from local conserved currents
\be
J_{\mu} \equiv W^{-1} {\tilde B}_{\mu} W \qquad \ra \qquad \pa^{\mu}
J_{\mu}=0
\label{conscurr}
\ee
The number of conserved currents is therefore equal to the dimension of the
abelian ideal where $B$ lives. 


Notice that the equations (\ref{localzc}) are sufficient local
conditions for the vanishing of the curvature (\ref{genzc}), 
in any generalized loop of $\Omega^{(d-1)}$. However,
if one is not interested in conserved charges one can restrict to the
case of infinitesimal loops only. In such case there are weaker local
sufficient conditions for the vanishing of the curvature
(\ref{genzc}). In the $2+1$ dimensional case for instance, those
conditions are given by
\be
\omega =0 \qquad \qquad \nu\(B\)=0
\label{weakerlzc}
\ee
with
\be
\omega = d B + A \wedge B \qquad \qquad \nu\(\cdot\) = 
\sbr{d A + A \wedge A - B}{\cdot}
\ee
In addition, one gets the Bianchi identities
\be
d\nu\(\cdot\) + A \wedge \nu\(\cdot\) = - \sbr{\omega}{\cdot}
\qquad \qquad 
d \omega + A \wedge \omega = \nu\(B\)
\ee
The two-form $\nu$ is called the fake curvature in the mathematical literature
 \cite{breen}. Of course, (\ref{localzc}) together with the fact $B$
 is abelian imply (\ref{weakerlzc}).

\section{The $\lambda$ -- $\bar{\lambda}$ realization of $sl(2)$}
\label{sec:lambda}

We now consider field theories such that the
equations of motion are written as the local zero curvature conditions
(\ref{localzc}). Since, the potential $B$ has to live on an abelian
ideal we consider a Poincar\'e type non-semisimple Lie algebra ${\cal
  G}=T+P$, satisfying
\be 
\sbr{T}{T} \subset T \qquad \sbr{T}{P} \subset P \qquad 
\sbr{P}{P} =0
\ee
where $T$ is a Lie algebra and $P$ a representation of it. In order
for the model to be integrable we need an infinite number of conserved
currents of the form (\ref{conscurr}), and consequently  the representation
$P$ has to be infinite dimensional.  In order to get that, we shall use the
Schwinger's construction. Let $R\(T\)$ be a (finite) matrix
representation of the algebra $T$, i.e.
\be
\sbr{R\(T\)}{R\(T^{\pr}\)} = R\(\sbr{T}{T^{\pr}}\)
\ee
Consider a number of oscillators equal to the dimension of $R$, 
\be
\sbr{a_i}{a_j}=0 \qquad \sbr{a_i^{\dagger}}{a_j^{\dagger}}=0 \qquad
\sbr{a_i}{a_j^{\dagger}}= \d_{ij} \qquad \quad i,j=1,\ldots {\rm
  dim}\; R
\ee
Then, it follows that the operators 
\be
S\( T\) \equiv \sum_{i,j} a^{\dagger}_i R_{ij}\(T\) a_j
\ee
constitute a representation of $T$ 
\be
\sbr{S\(T\)}{S\(T^{\pr}\)} = S\(\sbr{T}{T^{\pr}}\)
\ee 
The oscillators can be realized in terms of differential operators on
some parameters $\l_i$, as
$a_i \equiv \frac{\pa\;}{\pa \l_i}$ and $a_i^{\dagger} = \l_i$. 

In the case of the $sl(2)$ algebra, namely
\begin{equation}
[T_{3},T_{\pm}] = \pm T_{\pm} \; ; \qquad
[T_{+},T_{-}] = 2 T_{3}
\end{equation}
one gets that its two dimensional matrix representation leads to the
following realization in terms of differential operators
\begin{equation}
T_{+}\equiv \lambda \, \frac{d\;}{d\bar{\lambda}} \; ; \qquad
T_{-}\equiv \bar{\lambda} \, \frac{d\;}{d\lambda} \; ; \qquad
T_{3}\equiv \h\left( \lambda \, \frac{d\;}{d\lambda} - \bar{\lambda} \, \frac{d\;}{d\bar{\lambda}}\right)
\label{twoparrep}
\end{equation}

The states of the representations corresponding to such realization are
functions of $\lambda$ and $\bar{\lambda}$. 
The action of the operators are given by
\begin{eqnarray}
T_{3} \lambda^p\, \bar{\lambda}^q &=& \frac{p-q}{2} \,  \lambda^p\,
\bar{\lambda}^q \nonumber\\ 
T_{+} \lambda^p\, \bar{\lambda}^q &=& q \,  \lambda^{p+1}\,
\bar{\lambda}^{q-1} \nonumber\\ 
T_{-} \lambda^p\, \bar{\lambda}^q &=& p \,  \lambda^{p-1}\, \bar{\lambda}^{q+1}
\label{actiononl}
\end{eqnarray}
Notice that  from (\ref{actiononl}) the action of $T_3, T_\pm$ leaves the
sum of the powers of $\l$ and ${\bar \l}$ invariant.
Therefore, one can construct irreducible representations by
considering the states  
\begin{equation}
\ket{\left(p,q\right),m}\equiv \lambda^{p+m}\, \bar{\lambda}^{q-m}
\end{equation}
with $ m\in \IZ$ and $\left(p,q\right)$ being any pair of numbers
(real or even complex). 
Then
\begin{eqnarray}
T_{3} \ket{\left(p,q\right),m} &=& \left( \frac{p-q}{2} +m\right) \,
\ket{\left(p,q\right),m}\nonumber\\  
T_{+} \ket{\left(p,q\right),m} &=& \left( q-m\right) \,
\ket{\left(p,q\right),m+1}\nonumber\\ 
T_{-} \ket{\left(p,q\right),m} &=& \left( p+m\right) \,
\ket{\left(p,q\right),m-1} 
\label{actiononket}
\end{eqnarray}
On the subspace with fixed $(p+q)$, the Casimir operator acts as:
$$
\left(T_3^2 +{1\over 2} ( T_+ T_- + T_- T_+ ) \right)
\ket{\left(p,q\right),m} =  
s(s+1) \ket{\left(p,q\right),m},\quad
s = {1\over 2}(p+q)
$$
The parameter $s$ is the spin of the representation.

From the relations (\ref{actiononket}) one notices that if $p$ is an
integer, then $\ket{\left(p,q\right),-p}$ is a lowest weight state, since it
is annihilated by $T_{-}$. Analogously, if $q$ is an integer, then
$\ket{\left(p,q\right),q}$ is a highest weight state. If $p$ and $q$
are integers and $q>-p$, then the irrep. is finite dimensional. In
order to have integer spin representations we need $\frac{p-q}{2} \in
\IZ$. The spin zero state is $\ket{\left(p,q\right),-\frac{p-q}{2}} =
\left(\lambda\, \bar{\lambda}\right)^{\frac{p+q}{2}}$.  
Notice however, that not all irreps. for integer spin will have the zero spin
state. The reason is that if $p$ is a negative integer, or $q$ is a positive
integer, then  the
representation will truncate before reaching the zero spin state.

\section{The local zero curvature for theories with two scalar fields}
\label{sec:localzcsu2}

We shall consider field theories in a $d+1$-dimensional space-time and two
dimensional target space. So, the fields are Lorentz scalars  taking
values on the sphere 
$S^2$, torus $T^2$, the plane $R^2$, etc. We shall parametrize them by
a complex scalar field $u$. We introduce the
equations of motion through the local zero curvature conditions
(\ref{localzc}). We take $A_{\mu}$ to be a flat connection as in
(\ref{puregauge}) with the group element $W$ being given in the two
dimensional representation by 
\br
W = \frac{1}{\sqrt{1+\u2}} \; 
\pmatrix{1&iu\cr iu^* &1}
\er
Such element can be written in any representation as 
\begin{equation}
W \equiv e^{i u T_+}\, e^{\vp T_3}\, e^{i u^* T_-} \; ; \qquad 
\quad \quad 
\vp \equiv \log \left( 1 + \u2\right)
\label{w1w2bis}
\end{equation}
Then, using the realization (\ref{twoparrep}) one gets
\be
A_{\mu} \equiv \frac{1}{1+\u2}\left( -i \pa_{\mu} u \lambda 
\frac{d\;}{d\bar{\lambda}} 
-i \pa_{\mu} u^* \bar{\lambda} \frac{d\;}{d\lambda} 
+\left(u\pa_{\mu} u^* - u^*\pa_{\mu} u\right) 
\h\left( \lambda \, \frac{d\;}{d\lambda} - \bar{\lambda} \,
\frac{d\;}{d\bar{\lambda}}\right) \right) 
\label{s2pota}
\ee
The equations of motions come from the first relation in
(\ref{localzc}) (or equivalently (\ref{dualeqforb})), with the dual
potential (\ref{dualb}) being given by 
\be
{\tilde B}_{\mu}^{(s)} \equiv \frac{1}{1+\u2}\left(  \ck_{\mu} 
\lambda^{s+1} \bar{\lambda}^{s-1} -
 \ck_{\mu}^* \lambda^{s-1} \bar{\lambda}^{s+1}\right)
\label{s2potb}
\ee
The vector ${\cal K}_{\mu}$ is a priori a functional of $u$, $u^*$ and
their derivatives. Notice that we have chosen
$B_{\mu}^{(s)}$ to live in 
a representation where  $p=q=s$, and it has components only in the
direction of the states with eigenvalues $\pm 1$ of $T_3$. The factor 
$1/\(1+\u2\)$ was introduced for convenience since it could be absorbed
into the definition of ${\cal K}_{\mu}$.

Plugging (\ref{s2pota}) and (\ref{s2potb}) into (\ref{dualeqforb}) one
gets terms in the direction of states with with spins (eigenvalues of
$T_3$) $0$, $\pm
1$ and $\pm 2$. The terms with spins $-1$ and $-2$ are complex conjugate of
those with spins $1$ and $2$ respectively. The equations we get are
those given in (\ref{eqofmotion}), 
together with the complex conjugate of the first two equations. We
stress that these equations ensures the vanishing of the zero
curvature conditions (\ref{dualeqforb}) for any value of $s$.  
For the case $s=1$ the states with spins $\pm
2$ do not appear, and we do not have the second equation in
(\ref{eqofmotion}).  

The first two equations in (\ref{eqofmotion}) can be obtained through
another zero curvature equation. Consider the potentials \cite{AFSG97} 
\be
A_{\mu}^{(\a)} = \pa_{\mu} \a\(u\) Q \qquad \qquad 
{\tilde B}_{\mu}^{(\b)} = \b\(u\) {\cal K}_{\mu} P 
\label{heisenbergpot}
\ee
where $\a$ and $\b$ are functionals of $u$ only, and $Q$ and $P$
satisfy the Heisenberg algebra
\be
\sbr{Q}{P}=\one
\ee
The potential $A_{\mu}^{(\a)} $ is clearly flat,
and ${\tilde B}_{\mu}^{(\b)}$ does lie in an abelian ideal, generated
by $P$ and $\one$. Then the zero curvature condition
(\ref{dualeqforb}) gives
\be
D^{\mu} {\tilde B}_{\mu}^{(\b)} = \( \frac{\d \b}{\d u} \pa^{\mu} u \, 
{\cal K}_{\mu} + \b \,\pa^{\mu}{\cal K}_{\mu}  \) P 
+ \b\, \frac{\d \a}{\d u} \pa^{\mu} u {\cal K}_{\mu}\, \one
\ee
Therefore, its vanishing implies the first two equations in
(\ref{eqofmotion}). Replacing in (\ref{heisenbergpot}), ${\cal
  K}_{\mu}$ by its complex conjugate, and $\a$ and $\b$ by functions
of $u^*$ only, one gets through (\ref{dualeqforb}) the corresponding complex
conjugate equations. 

Notice that we get more equations than the number of fields, and so we have
a constrained system. However, there is a large class of vectors
${\cal K}_{\mu}$ for which the constraints are automatically
satisfied. Consider the vector 
\begin{equation}
K_{\mu} = h_{\mu\nu}\partial^\nu u =  \left( \partial^{\nu} u^* \partial_{\nu}
  u \right)  \, 
\partial_{\mu} u -  
\left(\partial_{\nu} u \right)^2 \, \partial_{\mu} u^*
\label{kmu}
\end{equation}
with $h_{\mu\nu}$ defined in (\ref{hmunudef}). 

Then, for any {\em real} function ${\cal F}(u,u^*, \partial u, \partial u^*)$
\begin{equation}
{\cal K}_\mu = {\cal F} K_\mu
\label{calK}
\end{equation}
satisfies identically the last two equations in (\ref{eqofmotion}). 

Therefore, for the class of vectors (\ref{calK}) the equations
(\ref{eqofmotion}) reduce to the single complex equation  
\begin{equation}
\partial^\mu ({\cal F} K_\mu) = 0
\label{model}
\end{equation}

Since (\ref{model}) guarantees the zero curvature (\ref{dualeqforb})
for any value of the spin $s$ in the potential (\ref{s2potb}), and for
any functional $\b\(u\)$ in (\ref{heisenbergpot}), it follows the
model possesses an 
infinite number of conserved currents of the form
(\ref{conscurr}). Notice that in the case of the potential
(\ref{s2potb}) the infinity of the currents come from the infinite
dimensional character of the abelian ideal, and in the case of
(\ref{heisenbergpot}) from the infinity of functionals $\b\(u\)$. We
will analyse these currents further in section \ref{sec:currents}.

\section{Existence of Lagrangean}
\label{sec:lagrangean}

The question then arises to know what are the models eq.(\ref{model})
which can be derived from an action principle
$$
S = \int d^{d+1} x {\cal L}
$$
Quite generally ${\cal L} $  is a function which is Lorentz invariant and depends only on first order derivatives. So it is of the form
$$
{\cal L} = {\cal L}( \sigma, \sigma^*, \rho,u,u^*)
$$
where we define
$$
\sigma = \partial^\mu u \partial_\mu u,\quad \sigma^* = \partial^\mu u^* \partial_\mu u^*,
\quad \rho = \partial^\mu u \partial_\mu u^*
$$
Then the equations of motion read
\begin{eqnarray*}
{\delta S \over \delta u } &=& \partial^\mu \left(
-2 {\partial {\cal L} \over \partial \sigma} \partial_\mu u - 
{\partial {\cal L} \over \partial \rho} \partial_\mu u^* \right) +  
{\partial {\cal L} \over \partial u } = 0 
\end{eqnarray*}
These equations of motion should be a linear combination of eq.(\ref{model})
and its complex conjugate. Hence,  we want to achieve
\begin{equation}
\partial^\mu \left(
-2 {\partial {\cal L} \over \partial \sigma} \partial_\mu u - 
{\partial {\cal L} \over \partial \rho} \partial_\mu u^* \right) +  
{\partial {\cal L} \over \partial u } = \Lambda \partial^\mu ({\cal F}K_\mu)
+ \Phi \partial^\mu ({\cal F}K^*_\mu)
\label{wewant}
\end{equation}
where  ${\cal L}$, ${\cal F}$, $\Lambda$ and $\Phi$ are functions to be determined.
Moreover,  ${\cal L}$ and ${\cal F}$ should be real.
Denote
$$
T_{\mu\nu} = \partial_\mu u \partial_\nu u, \quad 
T^*_{\mu\nu} = \partial_\mu u^* \partial_\nu u^*, \quad
S_{\mu\nu} = \partial_\mu u \partial_\nu u^* + \partial_\mu u^* \partial_\nu u
$$
The left hand side of eq.(\ref{wewant}) becomes
\begin{eqnarray*}
&&\left\{ -4{\partial^2 {\cal L} \over \partial \sigma^2} T_{\mu\nu}
-{\partial^2 {\cal L} \over \partial \rho^2} T^*_{\mu\nu}
-2 {\partial^2 {\cal L} \over \partial \sigma \partial \rho} S_{\mu\nu}
-2 {\partial {\cal L} \over \partial \sigma}\eta_{\mu\nu} \right\} \partial^{\mu\nu}u + \\
&& \left\{ -2{\partial^2 {\cal L} \over \partial \sigma \partial \rho} T_{\mu\nu}
-2{\partial^2 {\cal L} \over \partial \sigma^* \partial \rho} T^*_{\mu\nu}
-\left(2{\partial^2 {\cal L} \over \partial \sigma \partial \sigma^*} 
+ {1\over 2}{\partial^2 {\cal L} \over \partial \rho^2} \right) S_{\mu\nu}
- {\partial {\cal L} \over \partial \rho}\eta_{\mu\nu} \right\} \partial^{\mu\nu}u^* + \\
&&-2 \sigma {\partial^2 {\cal L} \over \partial u \partial \sigma} 
-2 \rho {\partial^2 {\cal L} \over \partial u^* \partial \sigma}  
- \rho {\partial^2 {\cal L} \over \partial u \partial \rho} 
- \sigma^* {\partial^2 {\cal L} \over \partial u^* \partial \rho} 
+ {\partial {\cal L} \over \partial u}
\end{eqnarray*}
while the right hand side comes in two pieces
\begin{eqnarray*}
&&\Lambda \left\{ 2\rho {\partial {\cal F} \over \partial \sigma} T_{\mu\nu} 
- \sigma {\partial {\cal F} \over \partial \rho} T^*_{\mu\nu} 
- \left(\sigma  {\partial {\cal F} \over \partial  \sigma}  
- {\rho \over 2}{\partial {\cal F} \over \partial \rho} + {1\over 2} {\cal F} \right)S_{\mu\nu} 
 + \rho {\cal F} \eta_{\mu\nu}  \right\} \partial^{\mu\nu} u+\\
&&\Lambda \left\{  \left( \rho {\partial {\cal F} \over \partial \rho} + {\cal F} \right)T_{\mu\nu}
- 2 \sigma  {\partial {\cal F} \over \partial  \sigma^*} T^*_{\mu\nu} 
+\left(\rho {\partial {\cal F} \over \partial \sigma^*} 
- {\sigma\over 2} {\partial {\cal F} \over \partial \rho} \right)S_{\mu\nu} 
 - \sigma {\cal F} \eta_{\mu\nu}  \right\} \partial^{\mu\nu}u^*  + \\
&& \Lambda  \left\{ {\partial {\cal F} \over \partial u^*}(\rho^2-\sigma\sigma^*) \right\}
\end{eqnarray*}
and
\begin{eqnarray*}
&&\Phi \left\{ - 2\sigma ^* {\partial {\cal F} \over \partial  \sigma} T_{\mu\nu} 
+\left( \rho{\partial {\cal F} \over \partial \rho}  + {\cal F} \right)T^*_{\mu\nu}
 + \left(\rho {\partial {\cal F} \over \partial \sigma} 
- {\sigma ^* \over 2}{\partial {\cal F} \over \partial \rho}\right) S_{\mu\nu} 
 - \sigma^* {\cal F^*} \eta_{\mu\nu}  \right\} \partial^{\mu\nu} u+\\
&&\Phi \left\{ - \sigma^* {\partial {\cal F^*} \over \partial \rho} T_{\mu\nu} 
+2\rho {\partial {\cal F} \over \partial \sigma^*} T^*_{\mu\nu} 
- \left( \sigma ^* {\partial {\cal F} \over \partial  \sigma^*}
- {\rho \over 2}{\partial {\cal F} \over \partial \rho} + {1\over 2} {\cal F} \right) S_{\mu\nu} 
 +\rho {\cal F} \eta_{\mu\nu}  \right\} \partial^{\mu\nu}u^*  + \\
&& \Phi  \left\{ {\partial {\cal F} \over \partial u}(\rho^2-\sigma\sigma^*) \right\}
\end{eqnarray*}

Identification of  the $\eta_{\mu\nu}$ terms gives:
\begin{eqnarray}
\Lambda {\cal F} = {\alpha \over X},&\quad& \alpha  \equiv  \sigma^* {\partial {\cal L} \over \partial \rho} 
+ 2 \rho {\partial {\cal L} \over \partial \sigma }  \label{lambdaF} \\
\Phi {\cal F} = {\beta \over X},&\quad& \beta  \equiv  \rho {\partial {\cal L} \over \partial \rho} 
+ 2 \sigma {\partial {\cal L} \over \partial \sigma } \label{lambdaF*} 
\end{eqnarray}
where we set $X=\sigma\sigma^*-\rho^2$.
Note that
$$
\rho \alpha  - \sigma^*\beta  = -2X {\partial {\cal L} \over \partial \sigma},
\quad -\sigma \alpha  + \rho \beta  = - X {\partial {\cal L} \over \partial \rho}
$$
The functions $\Lambda$ and $\Phi$ can now be eliminated from the equations. For instance
$\Lambda {\partial {\cal F} \over \partial \rho} = 
{\alpha \over X}{\partial \log {\cal F} \over \partial \rho} $ etc...
From the  $T^*_{\mu\nu}\partial^{\mu\nu}u $ term,
we extract $\partial \log {\cal F}/\partial \rho$: 
\begin{eqnarray}
 - X {\partial {\cal L} \over \partial \rho} {\partial \log {\cal F}\over \partial \rho }
&=& -X {\partial^2 {\cal L} \over \partial \rho^2}  - \beta  \label{FW2}
\end{eqnarray}
Similarly, from  the
$S_{\mu\nu}\partial^{\mu\nu}u $ term
we obtain $\partial \log {\cal F}/\partial \sigma$:
\begin{eqnarray}
 - X {\partial {\cal L} \over \partial \rho}{\partial \log {\cal F}\over \partial \sigma } 
&=& -X {\partial^2 {\cal L} \over \partial \sigma \partial \rho} 
+ \alpha  \label{FV2}
\end{eqnarray}
Finally from  the $T^*_{\mu\nu}\partial^{\mu\nu}u^* $ term,
we get  $\partial \log {\cal F}/\partial \sigma^*$:
\begin{eqnarray}
 - X {\partial {\cal L} \over \partial \rho}{\partial \log {\cal F}\over \partial \sigma^* } 
&=& -X {\partial^2 {\cal L} \over \partial \sigma^* \partial \rho} 
\label{FV*2}
\end{eqnarray}
Comparing eq.(\ref{FW2}) and its
complex conjugate, we find 
$ \beta ^* = \beta  $ which from the definition of $\beta $ implies ${\cal L} = {\cal L}(\sigma\sigma^*,\rho,u,u^*)$.
Similarly, comparing eq.(\ref{FV2}) with the complex conjugate of 
eq.(\ref{FV*2}) we find $\alpha =0$, hence $\Lambda = 0$ , which means 
${\cal L} = {\cal L}(X,u,u^*)$.
With this information at hand, we find
that the equations containing second order derivatives of $u$ reduce to
\begin{eqnarray*}
{\partial {\cal L} \over \partial X} {\partial \log {\cal F}\over \partial \rho } 
&=& -2 \rho {\partial^2 {\cal L} \over \partial X^2} \\
{\partial {\cal L} \over \partial X} {\partial \log {\cal F}\over \partial \sigma } 
&=& \sigma^* {\partial^2 {\cal L} \over \partial X^2} \\
{\partial {\cal L} \over \partial X} {\partial \log {\cal F}\over \partial \sigma^* } 
&=& \sigma {\partial^2 {\cal L} \over \partial X^2} 
\end{eqnarray*}
so that
$$
{\cal F} = f(u,u^*) {\partial {\cal L} \over \partial X},\quad 
\Longrightarrow \quad \Phi = {2\over f}
$$
with $f(u,u^*)$ real. Finally eq.(\ref{wewant}) reduces to
$$
{\partial {\cal L} \over \partial u } = -2X {\partial {\cal L} \over \partial X}
{\partial \log f  \over \partial u}
$$
which means that 
$$
{\cal L}(X,u,u^*) = {\cal L}\left( {X\over f^2(u,u^*) } \right)
$$
So  the Lagrangean is a function 
of the square of the pull back of an area form $H$ on a two
dimensional manifold 
$$
H = {1\over f} du \wedge du^*
$$

Let us summarize. Introduce 
$$
h_{\mu\nu} = 
\partial_{\mu} u \partial_{\nu} u^* - \partial_{\nu} u \partial_{\mu} u^* 
$$
The Lagrangean reads
$$
{\cal L} = {\cal L}\left({h^2 \over 2 f^2} \right),\quad h^2 = h_{\mu\nu}h^{\mu\nu}
$$
The equations of motion are
\be
\partial^\mu \left( {1\over f} {\cal L}' K_\mu \right) =0, \qquad\qquad
K_\mu = h_{\mu\nu} \partial^\nu u
\label{lagrangeanfinal}
\ee
or in expanded form:
$$
{1\over 2 f^3} {\cal L}'' K_\mu \partial^\mu h^2 
+ {1\over  f} {\cal L}' \partial^\mu K_\mu -
 {1\over  f^2}\left( {h^2\over f^2} {\cal L}'' + 
 {\cal L}' \right) K_\mu \partial^\mu f =0
$$
Comparing (\ref{lagrangeanfinal}) with (\ref{model}) we see that for the
models possessing Lagrangean we have ${\cal F}\equiv {\cal L}'/f$. 
Therefore, the effect of imposing the existence  of a
Lagrangean restricts the classes of fuctionals ${\cal F}$ to those
that depend on the derivatives of $u$ and $u^*$, only through
$h^2$. The dependency on the fields is still quite general.    

A special class of models models in a space-time of $d+1$ dimensions,
is obtained by choosing 
\begin{equation}
{\cal L} = \left({h^2 \over 2 f^2} \right)^{d/4}
\label{nicelagrangian}
\end{equation}
This has the advantage of circumventing Derrick's theorem, since the
energy of static configurations will be invariant under scale
transformations, $x^{\mu} \ra \l x^{\mu}$ \cite{AFZ992,kundu}.    
The equations of motion read
$$
f ( (d-4) K_\mu \partial^\mu h^2 + 4 h^2 \partial^\mu K_\mu ) 
-2 (d-2) h^2 K_\mu  \partial^\mu f = 0
$$
or
\begin{equation}
{\cal E} \equiv  (d-4) h_{\mu\nu}\,\partial^\nu  u \,\partial^\mu h^2 
+ 4 h^2 \partial^\mu h_{\mu\nu} \, \partial^\nu u 
+ (d-2) (h^2)^2 \partial_{u^*}\log f =0
\label{nicegene}
\end{equation}

Notice that for $d=2$ the equations of motion (and also the Lagrangean)
do not depend upon $f$,
and so on the determinant of the metric in target space. For $d=4$ the
theory has the advantage of having a Lagrangean which is quadratic in time
derivatives. 

In order to have  finite energy solutions one needs the field $u$ to
go to a constant at spacial infinity. Therefore, the space  $\IR^d$ can
be compactified on $S^d$, and so the finite energy solutions provide
mappings from $S^d$ to the target space $U$, which are classified by
the homotopy classes $\pi_d(U)$. 

In the case where $U\equiv S^2$ we have $f = (1+uu^*)^2$
where $u$ and $u^*$ are the stereographic projection coordinates on $S^2$: 
\be
\vec{n} = {1 \over 1+uu^*}( u+u^*, -i(u-u^*), -1 + uu^* ), \qquad  \vec{n}^2 = 1, 
\qquad u = {n_x +
i n_y \over 1 - n_z } 
\label{stereographicproj}
\ee
Then we have that for $d=2$ and $d=3$ the homotopy groups are 
isomorphic to the integers under addition,
i.e. $\pi_2(S^2)=\pi_3(S^2)=\IZ$. In fact, for $d=2$ we have the
winding number of $S^2 \to S^2$, and for $d=3$ the Hopf invariant of
$S^3 \to S^2$. So, in such cases we can have solutions
with non trivial topological charges.

\section{Conserved currents}
\label{sec:currents}

Since our models have a zero curvature representation
(\ref{dualeqforb}) with the potential (\ref{s2potb}) being valid in any spin
$s$ representation, it follows they have an infinite number of
conserved currents of the type (\ref{conscurr}). For {\em any} $s$,
the current  
$$
J_\mu^{(s)} = W^{-1} {\tilde B}_\mu^{(s)} W 
$$
is conserved:
$$
\partial^\mu J_\mu^{(s)} = 0
$$
Looking at eq. (\ref{s2potb}), we see that
$$
{\tilde B}^{(s)}_\mu = (\lambda \bar{\lambda} )^{(s+1)} {\tilde B}^{(-1)}_\mu
$$
If we consider a general ${\tilde B}_\mu = \sum_s \b_s {\tilde B}^{(s)}_\mu$,
we have
$$
{\tilde B}_\mu = b(\lambda \bar{\lambda}){\tilde B}^{(-1)}_\mu
$$
where $b(z)= \sum_s \b_s z^{s+1}$ is essentially an arbitrary function.

Using (\ref{twoparrep}) and (\ref{w1w2bis}) one can check that 
\begin{eqnarray}
W^{-1} f\left( \lambda , \bar{\lambda}\right) W = f\left(
\frac{\lambda - i u^*  \bar{\lambda}}{\sqrt{1+\u2}}, 
\frac{\bar{\lambda} - i u \lambda}{\sqrt{1+\u2}}\right) \label{conjw1w2}
\end{eqnarray}
Notice they look like a Lorentz boost on the space $\left( \lambda ,
\bar{\lambda}\right)$ and with complex 
velocity $iu$. Indeed, the covering of the Lorentz group is
$SL\left(2,\IC\right)$.

At the level of currents, this means
$$
J_\mu = b\left({(\lambda - iu^* \bar{\lambda} ) (\bar{\lambda} - iu \lambda ) 
\over 1 + uu^* } \right) J_\mu^{(-1)}
$$
We can write this in the nice form \cite{AFZ992,fujii}:
\be
J_\mu =  {\cal K}_\mu {\delta G \over \delta u} - 
{\cal K}^*_\mu {\delta G \over \delta u*} 
\label{niceformcurr}
\ee
where
$$
G = i \int^{v(u,u^*)} {dv \over v^2} b(v),\quad
v(u,u^*) = {(\lambda - iu^* \bar{\lambda} ) (\bar{\lambda} - iu \lambda ) 
\over 1 + uu^* }
$$
For this, we have to check that
$$
{\delta G \over \delta u} = {1\over (\bar{\lambda} -iu\lambda)^2} b(v),
\quad 
{\delta G \over \delta u^*} = {1\over (\lambda -iu^*\bar{\lambda})^2} b(v)
$$
which follows easily from
$$
{\delta v \over \delta u}= -i{v^2\over (\bar{\lambda} -iu\lambda)^2},
\quad
{\delta v \over \delta u^*}= -i{v^2\over (\lambda -iu^*\bar{\lambda})^2}
$$
Notice the currents obtained in that way have the functional $G$
depending on $u$ and $u^*$. However, one can get ``chiral'' currents 
using the potential (\ref{heisenbergpot}). Indeed  the currents
(\ref{conscurr}) for that potential are 
$$
J_{\mu}^{(\b)} = e^{\a \, Q} \, {\tilde B}_{\mu}^{(\b)} \, 
e^{-\a \, Q} = \b\,{\cal K}_{\mu} \, P + \a \, \b\,{\cal K}_{\mu} \,\one  
$$
and so they are of the form (\ref{niceformcurr}) with $G$ being a
function of $u$ only. 

Define
$$
\pi = {\partial {\cal L} \over \partial \dot{u}} = {1\over f^2} {\cal
  L}' h_{0\mu} 
\partial^\mu u^* = {1\over f^2} {\cal L}' K_0^* = {1\over f} {\cal K}_0^*
$$
and impose the Poisson bracket
$$
\{ \pi(x,t), u(y,t) \} = \delta(x-y)
$$
Then the conserved charges associate to (\ref{niceformcurr}) become 
$$
Q_G = i \int d^3x f \Big( \pi^* {\delta G \over \delta u} -  
\pi {\delta G \over \delta u^*} \Big)
$$
It is straightfoward to show that
$$
\{ Q_F,Q_G \} = i Q_{\{ F,G \}_{\rm Target} }
$$
where
$$
\{ F,G \}_{\rm Target} = f \left( {\delta F \over \delta u } {\delta G \over \delta u^* } -
{\delta F \over \delta u^* } {\delta G \over \delta u } \right)
$$
When the target space is $S^2$, this Poisson bracket is just the
expression of Kirillov bracket 
$ \{ n_i ,n_j \}_{S^2} = \epsilon_{ijk} n_k$ in the stereographic coordinates.
From this we immediately get
\be
\{ Q_G,u \} =- i f \partial_{u^*} G
\label{areadifftransf}
\ee
The meaning of the conserved currents is now clear:
$Q_G$ generates area preserving diffeomorphisms \cite{FR00}.  Indeed, we are
considering Lagrangeans of the form    
$$
{\cal L} = {\cal L}(H_{\mu\nu} H^{\mu\nu} )
$$
where 
$$
H =  {1\over f} h_{\mu\nu} dx^\mu \wedge dx^\nu = {1\over f} du\wedge du^*
$$
The form $H$ is he pullback of an area form on target space. 
Therefore the action is invariant under area preserving
diffeomorphisms. The conserved 
currents are the associated Noether currents.

The charge $Q_G$ generates the correct area preserving transformation for $u$
and  
$u^*$, when $G$ is real and a function of both  $u$ and $u^*$. In the case of
the diffeomorphisms coming from complex chiral functions depending on $u$ or
$u^*$ only, the charge that generates the correct transformation
under the Poisson bracket is a linear combination of charges of
different chiralities, i.e.
$$
Q_G = i \int d^3x f \Big( \pi^* {\delta G(u) \over \delta u} -  
\pi {\delta G^*(u^*) \over \delta u^*} \Big)
$$

\section{Symmetries}

We now want to study the symmetries of the equations of motion, and
not necessarily of the Lagrangean, of the models defined in
(\ref{nicelagrangian}). Extra symmetries, besides the area preserving
diffeomorphisms, appear as we will show, due to the scaling properties
of such models. We  apply Sophus Lie theory as explained for instance
in the book  \cite{olver}. A careful analysis shows that the
space-time and target space symmetries split into disjoint commuting
sets. We shall then treat them separately.

\subsection{Space time symmetries.}
\label{sec:spacetimesym}

 To say that 
$ x^{\rho}\to x^{\rho} + \xi^{\rho}$  is a symmetry of a differential equation 
$$
{\cal E}(x,u(x), \partial_{\mu} u(x),\partial_{\mu\nu} u(x))=0
$$  
where $\partial_{\mu\nu} \equiv \partial_{\mu}\partial_{\nu}$, means
that if $ u(x)$ is a solution  
so is $ {\tilde u} (x) \equiv u(x - \xi)$, i.e.,
\be
{\cal E}(x,{\tilde u}(x), \partial_{\mu} {\tilde u}(x),
\partial_{\mu\nu} {\tilde u}(x))=0
\label{eqonutilde}
\ee
Now, 
$$
\partial_{\mu} {\tilde u}(x) =\(\partial_{\mu} u\) \(x - \xi\) 
- \partial_{\mu} \xi^{\rho}(x) (\partial_{\rho} u)(x-\xi)
$$
and
\br
\partial_{\mu\nu} {\tilde u}(x) &=& (\partial_{\mu\nu}  u)(x-\xi ) - 
\partial_\mu \partial_\nu \xi^\rho (x) (\partial_\rho u)(x-\xi ) \nonu\\
&-& \partial_\mu \xi^\rho(x) \;(\partial_\rho \partial_\nu u)(x-\xi ) 
- \partial_\nu \xi^\rho(x) \;(\partial_\rho \partial_\mu u )(x-\xi )
\er
Evaluating (\ref{eqonutilde}) at $x+\xi$ and keeping only linear terms
in $\xi$, one gets that the vector field
\be
V = \sum_\rho \xi^\rho \partial_\rho
\label{vectorfield}
\ee
generates   a symmetry of the equations of motion if 
\be
\delta {\cal E}(x,u, \partial_{\mu} u,\partial_{\mu\nu} u)= 
\Lambda_1 \, {\cal E} + \Lambda_2 \, {\cal E}^*
\label{vareqofmotion}
\ee
with
\begin{eqnarray*}
\delta x^{\rho} &=& \xi^{\rho}\\
\delta u &=& 0\\
 \delta \,\partial_\mu u &=& - \partial_\mu \xi^\nu \partial_\nu u  \\
\delta\, \partial_\mu\partial_\nu u &=& - (\partial_\mu \partial_\nu \xi^\rho ) \partial_\rho u 
- \partial_\mu \xi^\rho \;\partial_\rho \partial_\nu u 
- \partial_\nu \xi^\rho \;\partial_\rho \partial_\mu u 
\end{eqnarray*}
Notice that the r.h.s. of  (\ref{vareqofmotion}) is taking into
account the fact that we require the variation of the equation of
motion to vanish only on the solution set of the complex equation ${\cal E}=0$.

Using this rule, one can evaluate the variations of
all the quantities appearing in the equations of motion. We
get for instance
\begin{eqnarray*}
\delta \, h_{\mu\nu} &=& - \partial_\mu \xi^\rho h_{\rho\nu} -
\partial_\nu \xi^\rho h_{\mu\rho} \\ 
\delta \,(\partial_\sigma h_{\mu\nu} ) &=&
- (\partial_\sigma \partial_\mu \xi^\rho) h_{\rho\nu} - 
(\partial_\sigma \partial_\nu \xi^\rho) h_{\mu\rho}
- \partial_\sigma \xi^\rho \partial_\rho h_{\mu\nu} - \partial_\mu
\xi^\rho \partial_\sigma h_{\rho\nu} 
- \partial_\nu\xi^\rho \partial_\sigma h_{\mu\rho}
\end{eqnarray*}
It follows that
\begin{equation}
\delta h^2 = -2 (\partial^\mu \xi^\rho + \partial^\rho \xi^\mu)
h_{\rho\nu} h_\mu ^{~~\nu} 
\label{deltah2}
\end{equation}
To have any chance for an invariance of the equations of motion, we require:
\begin{equation}
\partial^\mu \xi^\nu + \partial^\nu \xi^\mu = 2 D \eta^{\mu\nu}_{\rm eff.}
\label{spacesymmetry}
\end{equation}
where $D$ is the common value of $\partial_\mu \xi^\mu$ (no
summation). Notice that we have added a subscript {\em eff.} to the
metric. The reason for it is that we may consider  solutions of the
equations of
motion (\ref{nicegene}), which depend explicitly on the dimension
$d+1$ of space-time, 
but do not depend on some of the coordinates 
of space-time. That may
happen for instance when one considers  static
solutions. In the above construction only the effective coordinates
appear and we call the  dimension and metric of that subspace as
$d_{\rm eff.}$ and 
$\eta^{\mu\nu}_{\rm eff.}$ respectively. 
Then, we have
\begin{equation}
\delta \; h^2 = -4 D h^2
\label{deltah2b}
\end{equation}
and
$$
\delta \partial^\mu h_{\mu\nu} = -(\partial^2 \xi^\rho) h_{\rho \nu} 
-2 (\partial^\mu D ) h_{\mu\nu} - 2D \partial^\mu h_{\mu\nu}
- \partial_\nu \xi^\rho \partial^\mu h_{\mu\rho}
$$
but now
$$
\partial^2 \xi^\rho = \partial_\mu (\partial^\mu \xi^\rho) =
- \partial_\mu \partial^\rho \xi^\mu + 2 \partial_\mu D \eta^{\mu\rho}
= - \partial^\rho \partial_\mu \xi^\mu + 2\partial^\rho D 
= -(d_{\rm eff.}-2) \partial^\rho D 
$$
hence
$$
\delta \partial^\mu h_{\mu\nu} =(d_{\rm eff.}-4) \partial^\mu D h_{\mu\nu} 
-2D \partial^\mu h_{\mu\nu} - \partial_\nu \xi^\rho \partial^\mu h_{\mu\rho}
$$

From this, we quickly get
$$
\delta ( h^2 \partial^\mu h_{\mu\nu} \partial^\nu u) =
-8D \, h^2 \partial^\mu h_{\mu\nu} \partial^\nu u+ (d_{\rm eff.}-4) 
h^2 (\partial^\mu D) h_{\mu\nu} \partial^\nu u
$$
It is more cumbersome to obtain
$$
\delta ( h_{\mu\nu}  \partial^\nu u \,   \partial^\mu h^2 )=
-8D \, h_{\mu\nu}  \partial^\nu u \,   \partial^\mu h^2
- 4 h^2 (\partial^\mu D) h_{\mu\nu} \partial^\nu u
$$
So for ${\cal E}$ given by eq.(\ref{nicegene})
$$
\delta {\cal E} = -8D {\cal E} -4 (d-d_{\rm eff.}) h^2 \partial^\mu D
h_{\mu\nu} 
\partial^\nu u  
$$
Therefore, we conclude that
\begin{itemize}
\item If $d_{\rm eff.}=d$, e.g. for static solutions,
eq.(\ref{spacesymmetry})  
defines a symmetry of the equations of motion for all $D$.
\item If $d_{\rm eff.}\neq d$, eq.(\ref{spacesymmetry}) defines a
symmetry only  if $\partial^\mu D =0$.
\end{itemize}

Eqs.(\ref{spacesymmetry}) are not difficult to analyse.
Consider it for $\mu \neq \nu$, and apply $\partial_\mu \partial_\nu$.
Using that $D$ is the common value of $\partial_\mu \xi^\mu$ (no summation),
we get that $(\partial^\mu \partial_\mu + \partial^\nu \partial_\nu )D = 0$
for all pair of different indices. This implies that 
$\partial_\mu^2 D = 0$ for  
all $\mu$. Hence $D$ is at most linear in each of the variables $x^\nu$,
and $\xi^\mu$ is at most quadratic. But then
$$
\partial_\mu \partial_\nu D = \partial_\mu \partial_\nu^2 \xi^\nu =
- \partial_\nu^3 \xi^\mu = 0,\quad  \mu \neq \nu
$$
Hence $\partial_\mu \partial_\nu D = 0, \forall \mu,\nu$. So  
$D$ is a linear function of the $x^\nu$.
Then, we check that the set of vector fields $V^{(\mu)}$ (\ref{vectorfield})
with components
\begin{equation}
\xi^{(\mu)\nu} = x^\mu x^\nu - {1\over 2} \eta^{\mu\nu}_{\rm eff.} x^2
\label{conformalsymmetry}
\end{equation}
do satisfy the equations with $D^{(\mu)}= x^\mu$. This exhausts the
solutions with $D$ strictly linear in $x^\mu$.
 If $D$ is a constant, the solution is
 $$
 \xi^\nu = x^\nu
 $$
 and corresponds to dilatations. Finally if $D=0$, we find the translations,
 rotations and Lorentz boosts (Poincar\'e group).
 When $d_{\rm eff.}\neq d$ the transformations
 eq.(\ref{conformalsymmetry}) are   excluded, and we are left only with
 Poincar\'e and dilatation symmetries. 

\subsection{Target space symmetries.}
\label{sec:targetspacesym}

This time the vector fields 
generating the symmetries act only on the
target space, and the space-time coordinates are left unchanged, so 
$$
\delta x^\mu = 0
$$
and
$$
V = \Phi(u,u^*)\partial_u +  \Phi^*(u,u^*)\partial_{u^*}
$$
Therefore, since  $\delta$ and $\partial_\mu$ commute, one has
\br
\delta u &=& \Phi(u,u^*) \nonu\\ 
\delta \partial_{\mu} u &=& \partial_{\mu}\Phi(u,u^*) = 
\partial_u \Phi \partial_{\mu} u + \partial_{u^*} \Phi \partial_{\mu} u^* 
\er
and similarly for higher derivatives of $u$. Then
$$
\delta h_{\mu\nu} = 
 ( \partial_u \Phi  + 
\partial_{u^*} \Phi^*  ) h_{\mu\nu}
$$
so that
$$
\delta h^2 = 2  ( \partial_u \Phi + 
\partial_{u^*} \Phi^*  ) h^2
$$
We find easily
\begin{eqnarray*}
\delta ( h_{\mu\nu} \partial^\nu u\partial^\mu h^2) &=&
 (4 \partial_u \Phi + 
3\partial_{u^*} \Phi^*  ) h_{\mu\nu} \partial^\nu u \partial^\mu h^2
+ \partial_{u^*} \Phi  h_{\mu\nu}\partial^\nu u^* \partial^\mu h^2 \\
&-&  \partial_{ u^*} ( \partial_u \Phi + 
\partial_{u^*} \Phi^*  ) (h^2)^2 \\
\delta ( h^2\partial^\mu h_{\mu\nu} \partial^\nu u) &=&
 (4 \partial_u \Phi + 
3\partial_{u^*} \Phi^*  )h^2\partial^\mu h_{\mu\nu} \partial^\nu u
+  \partial_{u^*} \Phi   h^2\partial^\mu h_{\mu\nu} \partial^\nu u^*\\
&-&{1\over 2} \partial_{ u^*} ( \partial_u \Phi + 
\partial_{u^*} \Phi^*  ) (h^2)^2 
\end{eqnarray*}
Hence, again for ${\cal E}$ given by eq.(\ref{nicegene}), we find
$$
\delta {\cal E} = (4 \partial_u \Phi + 
3\partial_{u^*} \Phi^* ) \;{\cal E}
- \partial_{u^*} \Phi  \;{\cal E}^*
-(d-2) Q \; (h^2)^2
$$
where
\br
Q &=& \partial_{ u^*}( \partial_u \Phi  + 
\partial _{u^*}\Phi^* ) -
\partial_{u} \log f \; \partial_{u^*}  \Phi  - 
\partial_{u^*} \log f  \; \partial_{u^*} \Phi^*\nonu\\
&&- \partial_{u}\partial_{u^*}\log f \; \Phi 
-\partial_{u^*}\partial_{u^*}\log f \;  \Phi^* 
\er
If $d=2$ we have a symmetry for any diffemorphism in target
space. If $d\neq 2$, we have a symmetry only if $Q=0$.
Setting, in that case, 
$$
\Phi = f \widetilde{\Phi}
$$
the condition $Q=0$ becomes
\begin{equation}
\partial_{u^*} ( f [  \partial_u \widetilde{\Phi}  + 
\partial _{u^*}\widetilde{\Phi}^* ] ) = 0
\label{targetsymmetry}
\end{equation} 
So, the quantity $f [  \partial_u \widetilde{\Phi}  + 
\partial _{u^*}\widetilde{\Phi}^*]$ depends on $u$ only, but since it
is real it must be a constant. We have to distinguish two cases. The
first happens  when
that constant vanishes, and one has 
\be
\partial_u \widetilde{\Phi}  + 
\partial _{u^*}\widetilde{\Phi}^*  = 0
\label{notextra}
\ee
Setting, for complex  $F$ : 
\be
\widetilde{\Phi} = \partial_{u^*} F 
\label{fdefphi}
\ee
the equation (\ref{notextra})  becomes
$$
 \partial_u \partial_{u^*}(  F + F^*) = 0
$$
Therefore, $F$ is either pure imaginary or $F$ is a function of
$u^*$ only. 
These cases correspond to  the area preserving
diffeomorphisms. 

The second case happens when the integration constant does not vanish
(which we choose  equals to $2$ for convenience), and so   
\begin{equation}
\partial_u \widetilde{\Phi}  + 
\partial _{u^*}\widetilde{\Phi}^*  = {2 \over f}
\label{niceextra}
\end{equation}
Setting  $F$ real in (\ref{fdefphi}) (since the imaginary part would
lead to the area 
preserving diffeomorphisms discussed in the first case above) 
the equation (\ref{niceextra}) becomes
$$
 \partial_u \partial_{u^*} F = {1\over f}
$$
The function $F$ generates
a new symmetry, not preserving the area:
$$
F =  \int^u du \int^{u^*} du^* {1\over f}
$$
Integration constants have to be chosen in order that $F(u,u^*)$ is real.

\bigskip

For the sphere, we have $f = (1+uu^*)^2$ and we find
\be
F = \log (1+ uu^*), \quad \Longrightarrow \Phi = (1+ uu^*) u
\label{extratransf}
\ee

For the plane, we have $f = 1$ and we find
$$
F = uu^*, \quad \Longrightarrow \Phi =  u
$$
and so, it is a scaling transformation on the target space (plane).

\section{Solutions.}

 Let us consider the equations of motion (\ref{nicegene}) in the
 static case. According to the analysis of section
 \ref{sec:spacetimesym} we have $d_{\rm eff.}=d$. The symmetries 
 eq.(\ref{conformalsymmetry}) read
\be
{x'}^i = x^i + (\epsilon \cdot x) x^i - {1\over 2} x^2 \epsilon^i
\qquad \qquad i=1,2,\ldots d
\label{staticconftrans}
\ee
They are conformal symmetries in the $d$-dimensional Euclidean space,
because ${dx'}^2 = (1 + 2 \epsilon \cdot x) dx^2$. 
These symmetries can be used to find static solutions to the equations
of motion (\ref{nicegene}).
 
To do that, first recall the description
 of conformal symmetries of Euclidean space. Consider points of Euclidean
 space as spheres of radius zero.
 The equation for spheres is of the form (we use arrows to denote
 vectors in $d$-dimensional Euclidean space)
 $$
 \alpha \vec{x}^2 - 2 \vec{\beta} \cdot \vec{x}  + \gamma = 0
 $$
or
$$
\Big( \vec{x} -{\vec{\beta} \over \alpha} \Big)^2 = {\vec{\beta}^2 -
\alpha\gamma \over \alpha^2} 
$$
So Euclidean space is described by 
\begin{equation}
\vec{\beta}^2 - \alpha\gamma = 0
\label{quadrics}
\end{equation}
and 
\begin{equation}
\vec{x} = {\vec{\beta }\over  \alpha}
\label{xdebeta}
\end{equation}
Any linear transformation over the parameters $\alpha$, $\vec{\beta}$
and $\gamma$, which preserves eq.(\ref{quadrics}), acts 
on $\vec{x}$ through eq.(\ref{xdebeta}).
These are the conformal transformations. Write the quadratic
expression (\ref{quadrics}) as
$$
\vec{\beta}^2 - \alpha\gamma  = \sum_{i=1}^d \beta_i^2 +
\Big({ a \alpha - a^{-1} \gamma \over 2} \Big)^2 -
\Big({ a \alpha + a^{-1} \gamma \over 2} \Big)^2
$$
where $a$ is an arbitrary scale. The group that leaves that quadractic
form invariant is  the
conformal group  $O(d+1,1)$. Denoting $d\equiv 2n$ or
$d=2n-1$, according to $d$ being even or odd, we see that the maximum number
of {\em commuting, compact rotations} is equal to $n$.   
In the case of $d$ odd, the first $n-1$ commuting rotations can be
taken  on the planes
$(\beta_{2j-1}, \beta_{2j})$, i.e. 
\be
\delta \beta_{2j-1} = \beta_{2j}, \quad 
\delta \beta_{2j} = -\beta_{2j-1} \qquad \qquad j=1,2,\ldots n-1 
\label{betarotation}
\ee
The other one is then the rotation $\Big( \beta_{2n-1}, { a \alpha - a^{-1}
\gamma \over 2}\Big)$: 
\be
\delta \beta_{2n-1} ={ a \alpha - a^{-1} \gamma \over 2}, \quad
\delta { a \alpha - a^{-1} \gamma \over 2} =-\beta_{2n-1}, \quad
\delta { a \alpha + a^{-1} \gamma \over 2} = 0
\label{alphabetarotation}
\ee
In the case of $d$ even, we have basically two distinct
possibilities. In the first one we take $n$ commuting rotations as in
(\ref{betarotation}), but with $j$ varying from $1$ to $n$. In the
second case we take the coordinate $\beta_{2n}$ out of the game, and
choose the $n$ commuting rotations as in the odd $d$ case,
i.e. rotations (\ref{betarotation}) and (\ref{alphabetarotation}). 

Notice that when the rotations (\ref{betarotation}) are transported to
$\vec{x}$-space using (\ref{xdebeta}) one gets rotations on the planes 
$(x_{2j-1}, x_{2j})$, i.e. 
\be
\delta x_{2j-1} = x_{2j}, \quad 
\delta x_{2j} = -x_{2j-1} \qquad \qquad j=1,2,\ldots n-1 
\label{xrotation}
\ee
 
Now, let us transport the rotation (\ref{alphabetarotation})  to 
$\vec{x}$-space. We have from  (\ref{alphabetarotation}) and
(\ref{quadrics})  
$$
\delta \beta_{2n-1} = {a\over 2} \alpha - {a^{-1}\over 2\alpha}\vec{\beta}^2,
\quad \delta\alpha =- a^{-1}\beta_{2n-1},\quad
\delta\gamma = a \beta_{2n-1}
$$
Of course, in the case of $d$ even the extra $\beta$ variable is left
unchanged, i.e. 
$$
\delta \beta_{2n}=0
$$
Therefore for both cases, $d$ even or odd, we have that
\begin{eqnarray*}
\delta x_i &=& a^{-1} {\beta_i \beta_{2n-1} \over \alpha^2} =  
{ x_i x_{2n-1} \over a} \qquad \qquad i\neq 2n-1\\
\delta x_{2n-1} &=&  {a\over 2} - 
{a^{-1}\over 2 } {\vec{\beta}^2 \over \alpha^2} + 
a^{-1} {\beta_{2n-1}^2 \over \alpha^2}
=  {1\over 2 a} (x_{2n-1}^2 - \sum_{i\neq 2n-1} x_i^2 ) + {a\over 2} 
\end{eqnarray*}
We recover exactly one of our conformal transformations 
 (\ref{staticconftrans}) with $\epsilon^i = \epsilon \delta_{i,2n-1}$, 
 plus a translation in the $x_{2n-1}$ direction. 

Define vector fields corresponding to 
the  rotations:
\begin{eqnarray}
{\partial_{\theta_i}  } &=& x_{2i} \partial_{x_{2i-1}} 
- x_{2i-1} \partial_{x_{2i}}
 \qquad \qquad i=1,2,\ldots  n-1 
\label{vecfield1}\\
{\partial_ {\theta_{n}} } &=& \frac{x_{2n-1}}{a} \sum_{i\neq 2n-1} x_i
\partial_{x_i} 
+{1 \over 2 a} ( a^2 +x_{2n-1}^2 - \sum_{i\neq 2n-1} x_i^2 )
\partial_{x_{2n-1}} 
\label{vecfield2} 
\end{eqnarray}
In the $u$ internal space we have a compact rotation which is the
phase transformations of $u$, i.e. $u \to u e^{i\varphi}$.  
According to S. Lie, educated 
 Ansatz for solving the equations of motion are obtained by imposing invariance under a combination of
internal and external rotations , i.e. of the 
commuting vector fields
\begin{eqnarray}
\relax [{\partial_{\theta_j}  } -im_j (u\partial_u - u^*\partial_{u^*}
) ] u 
= 0 \qquad \qquad j=1,2,\ldots  n
\label{sym1}
\end{eqnarray}
where $m_j$ are integer numbers. Then we have
$$
u = R(\zeta_l) e^{i\sum_{j=1}^n m_j\theta_j } 
$$
where $\zeta_l$ are such that 
$\partial_{\theta_j} \zeta_l = 0$, and are given as 
$$
\zeta_l = { a^2 (x^2_{2l-1} + x^2_{2l} ) \over 
(a^2 + r^2 )^2 } \qquad \qquad 
l=1,2,\ldots n-1
$$
where $r^2\equiv \sum_{i=1}^{d} x_i^2$. 
In the case of $d$ even, we have an extra $\zeta$ variable which is
$$
\zeta_n = { a^2 x^2_{2n}  \over 
(a^2 + r^2 )^2 } 
$$

For the other choice of commuting rotations in the case of $d$ even,
namely the ones in (\ref{betarotation}) but with $j$ varying
from $1$ to $n$,   the transformations in
$\vec{x}$-space  and corresponding vector fields are all of the form
(\ref{xrotation}) and (\ref{vecfield1}) respectively.  
Therefore, proceeding in the same manner we find the ansatz
$$
u = S(\rho_l) e^{i\sum_{j=1}^n m_j\theta_j } 
$$
where $\partial_{\theta_j} \rho_l=0$, i.e. 
$$
\rho_l = \frac{1}{a} \, \sqrt{x^2_{2l-1} + x^2_{2l}} \qquad \qquad 
l=1,2,\ldots n
$$

The $n$ {\em commuting} conditions eqs.(\ref{sym1}) introduce 
$n$ cyclic coordinates $\theta_j$, that will disappear from
the equations of motion\footnote{Remember that the variation of the
equations of motion under a given symmetry implies $\delta {\cal E} =
\partial_{\theta_j}{\cal E}= \Lambda_j {\cal E}$. However, since the
vector fields commute there must exist a $\chi$ such that
$\Lambda_j = \partial_{\theta_j} \chi$ Then, 
${\cal E}(\theta_i, \zeta_l) = \exp( \chi ) 
{\cal E}(0, \zeta_l)$.}. Hence, we know we will get a single differential
equation 
for $R(\zeta_l)$ (and for $S(\rho_l)$).
A priori $R(\zeta_l)$ and $S(\rho_l)$ are complex function. The form
of the Ansatz 
is preserved if we perform an area preserving diffeomorphism depending
only on $uu^*$. Indeed, from (\ref{areadifftransf})
$$
\delta u = i f \partial_{u^*} G(uu^*) = if G'(uu^*) u
$$
Therefore, if $f$ does not depend upon the phase of $u$, like in the
sphere where $f = (1+uu^*)^2$, or on the plane where $f=1$, then
change in $u$ is proportional to $u$ and the function 
multiplying it do not depend upon the phase of $u$. So, the form of the Ansatz
is indeed preserved and 
$$
\delta R = i (1+RR^*)^2G'(RR^*) R
$$
This implies that
$$
\delta (RR^*) = 0,\quad {\rm and} \quad
\delta \Big({R\over R^*}\Big) = 2i (1+RR^*)^2G'(RR^*) {R\over R^*}
$$
the real function $G$ can be used to kill one real function,
for instance the phase of $R$. Hence, we can assume 
$R$ to be real. The same arguments apply to $S(\rho_l)$.

Let us particularize this construction for the  
two lowest dimensions. In the case $d=2$ we have that the static model is
trivial. Indeed, its equation of motion  (\ref{lagrangeanfinal}) is
given by
$$
\partial^{\mu}\( \frac{h_{\mu\nu}}{\sqrt{h^2}} \, \partial^{\nu} u\) =0
$$
In the static case $h_{\mu\nu}$ has just one component and so the
above equation is satisfied by any configuration for the $u$ field.

In the case $d=3$  the coordinates $\zeta_1\equiv \zeta$, 
$\theta_{1}\equiv \varphi$, $\theta_{2}\equiv \xi$,
are related to toroidal coordinates. Recall that the toroidal
coordinates are defined as  
\begin{eqnarray*}
x &=& a q^{-1} \sinh \eta \cos \varphi \\
y &=& a q^{-1} \sinh \eta \sin \varphi \\
z &=& a q^{-1} \sin \xi   \\
q &=& \cosh \eta - \cos \xi
\end{eqnarray*}
Computing the vectors fields
$\partial_{\varphi}$ and $\partial_{\xi}$ one recovers exactly
(\ref{vecfield1}) and (\ref{vecfield2}) respectively. Moreover, 
we have 
\be
\zeta  = \frac{a^2(x^2+y^2)}{(a^2+x^2+y^2+z^2)^2}= {1\over 4} \tanh^2 \eta
\label{3dzeta}
\ee

In the toroidal coordinates, the equation we get for $R(\eta)$ reads
\cite{AFZ992} 
$$
{\partial \over \partial \eta} \log { RR'\over (1+R^2)^2 } = - { 2 m^2
\sinh^2 \eta - n^2 \cosh \eta  
\over m^2 \sinh^2 \eta + n^2 \sinh \eta }
$$
Imposing the boundary conditions (see (\ref{stereographicproj})) 
$$
\vec{n} \to (0,0,1) \quad {\rm or} \quad |u| \to \infty  \quad {\rm
or} \quad R \to \infty  \quad  
{\rm as} \quad  \eta \to 0
$$
$$
\vec{n} \to (0,0,-1) \quad {\rm or} \quad u \to 0  \quad {\rm or}
\quad R \to 0  \quad  
{\rm as} \quad  \eta \to \infty
$$
one gets
\begin{equation}
u = R(\eta) e^{i(m\xi + n \theta) }, \quad R^2 = { \cosh \eta - \sqrt{
n^2/m^2 + \sinh^2 \eta } 
\over \sqrt{1 + m^2/n^2 \sinh^2 \eta } - \cosh \eta }
\label{toroidal}
\end{equation} 
The Hopf charge and the energy are 
$$
Q_H = - nm, \quad E = (2\pi)^2  \sqrt{ |n||m|(|n|+|m|) }
$$
The origin of the ansatz proposed in \cite{AFZ992} to construct the
hopfion solutions given above, is now clear: it comes from the
conformal symmetry  of the equations of
motion in the $3$-dimensional space. The toroidal coordinates appear
 because the two commuting conformal transformations coincide 
to  rotations along the  angles $\varphi$ and $\xi$.

\section{Extra Noether current}

The extra target space symmetry of the equations of motion, found in
(\ref{extratransf}), is not a symmetry of the action. However,
combining it with the space-time scaling transformation, one can find a
symmetry of the action and so a Noether current. Although the
procedure works for all Lagrangeans of the type (\ref{nicelagrangian})
we will consider the $3+1$ dimensional model only ($d=3$), on the sphere. 

To compute the Noether currents,  we shall use
the old trick of making the parameter of the transformation to depend
upon the space-time coordinates. Suppose a given action is invariant
under a global transformation. If we make the parameter to depend on
$x^{\mu}$ we should have that its variation is of the form
$$
\delta S = \int d^4 x \, J^{\mu} \partial_{\mu} \vareps
$$
However, any variation of the action vanishes on shell.  Hence,
integrating  by parts we get  that the conservation law
$$
\partial_{\mu} J^{\mu}=0
$$
holds  when the equations of
motion are satisfied.

Under the target space transformation (\ref{extratransf}) we have that
$$
\delta \left( {h^2 \over 2 f^2} \right) = 2 \vareps f
(\partial_u \widetilde{\Phi}  + \partial_{u^*} \widetilde{\Phi}^* )
\left( {h^2 \over 2 f^2} \right) +
{2\over f} h^{\mu\nu} ( \widetilde{\Phi} \partial_\nu u^* 
- \widetilde{\Phi}^* \partial_\nu u) \partial_\mu\vareps
$$
For the area preserving diffeomorphisms the $\epsilon$ term vanishes
(see (\ref{notextra})). Then, one obtains the Noether
current 
$$
J^{\mu}_{\rm area~pres.}={3\over 2f}  \left( {h^2 \over 2 f^2} \right)^{-1/4}
 h^{\mu\nu} ( \widetilde{\Phi} \partial_\nu u^* 
- \widetilde{\Phi}^* \partial_\nu u)
$$
which correspond to the currents (\ref{niceformcurr}). 

In the case of the extra transformation (\ref{niceextra}) we get
$$
\delta \left( { h^2 \over 2 f^2 } \right) = 4 \vareps\, \left( { h^2
\over 2 f^2 } \right)  +
{2\over f} h^{\mu\nu} ( \widetilde{\Phi} \partial_\nu u^* 
- \widetilde{\Phi}^* \partial_\nu u) \partial_\mu\vareps
$$
and so, for global transformation the action is not invariant, since 
$\delta S = 3\vareps\,  S$. 

To compensate for that non invariance we consider space-time scaling
transformation of the form ${x'}^{\mu } = x^\mu -3 \vareps\, x^\mu $,
where the factor $-3$ was introduced to compensate the factor $3$
above. We have that
 \be
 \d\(d^4x\) = -12 \, \vareps \, d^4x - 3 x^{\nu}\,\pa_{\nu}\vareps \, d^4x
 \ee
 Moreover from eq.(\ref{deltah2}) (or (\ref{deltah2b}) with $\xi^\mu =
 -3\vareps x^\mu$,  and so $D=-3$), we
 get:
 $$
 \delta h^2 = 12 \vareps h^2 +12 x^\rho h_{\rho\nu} h^{\mu\nu}
 \partial_\mu \vareps 
 $$
The variation of the action coming from 
space dilatations is:
$$
\delta S = -3 \vareps S 
+ \int d^4x \left( {h^2 \over 2 f^2} \right)^{-1/4} \( {9\over 2 f^2}  
x^\rho h_{\rho\nu} h^{\mu\nu} 
- 3 x^{\mu}\,   {h^2 \over 2 f^2}  \) \partial_\mu \vareps 
$$
Therefore, combining the two transformations as
\begin{equation}
\delta u = \vareps \Phi , \quad \delta x^\mu = -3\vareps x^\mu
\label{nicesym}
\end{equation}
we get a symmetry of the action, and the corresponding Noether current
is 
\br
 J^{\mu}\equiv - x^{\mu}{\cal L} 
 + 
  \left( {h^2 \over 2 f^2} \right)^{-1/4} {1\over 2f^2}
  \Big( h^{\mu\nu} ( \Phi \partial_\nu u^*  - \Phi^* \partial_\nu u)
 +3 x^\rho h_{\rho\nu} h^{\mu\nu} \Big)
 \label{nicecur}
 \er
Recall that $\Phi$ is given at the end of section
\ref{sec:targetspacesym} for the cases of the sphere and plane.

 There is a connection between this Noether current and the
  energy-momentum tensor. 
  The canonical energy-momentum tensor is defined by
 \begin{equation}
 \Theta_{\mu\nu} = \frac{\pa\cl}{\pa \pa^{\mu}u}\, \pa_{\nu} u + 
 \frac{\pa\cl}{\pa \pa^{\mu}u^*}\, \pa_{\nu} u^* - g_{\mu\nu} \cl
 \end{equation}
 and the conserved current (\ref{nicecur}) can be written as
 \be
 J_{\mu} = x^{\nu} \Theta_{\mu\nu} + j_{\mu} 
\label{currenergymom} 
\ee
 with
 \be
 j_{\mu} \equiv \left( {h^2 \over 2 f^2} \right)^{-1/4} {1\over 2f^2}
   h^{\mu\nu} ( \Phi \partial_\nu u^*  - \Phi^* \partial_\nu u)
 \ee

 We now introduce the charge
 \be
 Q = \int d^3x\; J^0 = \int d^3x\;\( x^{\nu} \Theta_{0\nu} + j_{0}\) 
 \ee
 Then, for static configurations  
 \be
 \frac{d Q}{dt} = - \int d^3x\; \cl = E \equiv \mbox{\rm static
 energy} =  \int d\Sigma_i J_i
 \ee

 From the above formula, the static energy of the toroidal solutions can be 
  written as the flux of the current $\vec{J}$ through some boundary surface 
surrounding the singularities of the current $\vec{J}$.   
The field $u$ becomes singular when 
  $\eta \to 0$: 
  $$ 
  u = C_{n,m} {1\over \eta } e^{im\xi + in \varphi } ,\qquad\qquad C_{n,m} =
  |n|^{3/2}\sqrt{2 \over | n | | m |(|m| + | n | ) } 
  $$ 
 From eq.(\ref{3dzeta}) we have, when $\eta$ is small 
 (we set $a=1$) 
 $$ 
 \eta = {2 \rho \over 1 + r^2 }, \quad q = 
 {2\over 1 + r^2},\quad  \sin \xi = {2z\over 1+r^2},\quad \tan \varphi = 
 {y\over x} 
 $$ 
 from what we see that $\eta$ is small either when $\rho$ is small
 ($z$-axis) or $r$   is large (sphere at infinity).  

One can check that the $x^{\nu}\Theta_{\mu\nu}$ part of
(\ref{currenergymom})  is regular, and
that the $j_{\mu}$ part is regular at infinity. On the $z$-axis it
behaves like  
$$
\vec{j}\cdot {\hat e}_{\rho} = 2\sqrt{| n | | m |(|m| + | n | )}\;
\frac{1}{1+z^2} \frac{1}{\rho}  \qquad \qquad \rho \sim 0  
$$
Therefore, the flux through an infinitesimally thin and infinite  tube
around the $z$-axis 
is obtained from the above expression by multiplying it by $\rho d\vp
dz$, and integrating over $\vp$ and $z$, 
to get 
$$
{\rm flux} \; \; = (2\pi)^2 \sqrt{| n | | m |(|m| + | n | )} = \; \;
{\rm static~energy}
$$
We observe that the hopfion solutions carry a line structure
with it, which is that line of singular flux of the conserved current
(\ref{currenergymom}).

\section{Conclusion.}

We have seen that the  ideas of \cite{AFSG97} to extend in higher
dimensions some notions of integrability, do provide us with 
models with an infinite number of conserved currents.
It turned out that  
these currents, in the case of the models discussed here,  
are  associated to an invariance of the 
Lagrangean under area preserving diffeomorphisms.
It is intriguing that they do not play an important role in the
construction of the static solutions of the equations of motion. 
That role is played by the
conformal symmetry of the static equations of motion, which leads to
the construction of the relevant ansatz. It remains to see the role of
the conserved currents  in the scatterring processes of
these solutions.

\vspace{1cm}

\noindent{\large {\bf Acknowledgements}}

We are greatful to M. Bellon for discussions. We acknowledge the
support of CAPES/COFECUB under the grant 306/00-II. LAF is partially
supported by CNPq.

\end{document}